# Multimodal Learned Sparse Retrieval for Image Suggestion

Thong Nguyen    Mariya Hendriksen    Andrew Yates
University of Amsterdam
Amsterdam, The Netherlands
{t.nguyen2,m.hendriksen,a.c.yates}@uva.nl

## ABSTRACT

Learned Sparse Retrieval (LSR) is a group of neural methods designed to encode queries and documents into sparse lexical vectors. These vectors can be efficiently indexed and retrieved using an inverted index. While LSR has shown promise in text retrieval, its potential in multi-modal retrieval remains largely unexplored. Motivated by this, in this work we explore the application of LSR in the multi-modal domain, i.e., we focus on Multi-Modal Learned Sparse Retrieval (MLSR). We conduct experiments using several MLSR model configurations and evaluate the performance on the image suggestion task. We find that solving the task solely based on the image content is challenging. Enriching the image content with its caption improves the model's performance significantly, implying the importance of image captions to provide fine-grained concepts and context information of images. Our approach presents a practical and effective solution for training LSR retrieval models in multi-modal settings.

## 1 INTRODUCTION

Learned Sparse Retrieval (LSR) is a neural retrieval method that encodes queries and documents to bags of tokens, which could be indexed and retrieved efficiently by an inverted index.

**Image Suggestion Task.** The task is defined as follows: given a query and a set of candidates, we rank all candidates w.r.t. their relevance to the query. The query is a text, whereas the set of candidate items are images. Hence, we aim to retrieve relevant images that accurately describe the textual query.

**Vision-Language Learned Sparse Retriever.** We propose to address the task using a Vision-Language Sparse Retriever approach. The architecture is a bi-encoder, including a query encoder and document encoder. Both query and document encoders are based on transformer encoder architecture with either a Multi-Layer Perceptron (MLP) or Masked Language Model (MLM) sparse projection head on top. We experiment with the following model configurations:

- $M_{\mathcal{T} \rightarrow \mathcal{I}}$: the model uses image information to build document representations.

- $M_{\mathcal{T} \rightarrow \mathcal{T}}$: the model uses textual information to build document representations.
- $M_{\mathcal{T} \rightarrow \mathcal{T}+\mathcal{I}}$: the model uses both textual and image information to build document representations.

## 2 APPROACH

We follow the same terminology and notation as in [13, 31]. The input dataset can be represented as image-text pairs. For the image suggestion task, we use textual data as a query and we aim to retrieve a ranked list of top-$k$ images that describe the textual query.

**Vision-Language Learned Sparse Retrieval.** To address the task, we explored the application of learned sparse retrieval that leverages visual and textual information. The model comprises a query and document encoder. Depending on the modality of the query and document, the encoder could either be an MLP and MLM encoder. The MLP encoder can only used with text modality, while the MLM is applicable to both textual and visual data. Note that in this task, the document could be either an image, caption, or both. Figure 1 illustrates the architecture of the MLP and MLM encoder.

**MLP Encoder.** The MLP encoder takes a text as input and produces an important weight for each token of the input text. For example, for the input text "*text image retrieval*", an MLP encoder outputs weights, such as *{"text": 10 , "image": 20 , "retrieval": 50 }*.

An MLP encoder is a network that takes a sequence of contextualized embeddings $h_j$ produced by the dense encoder for each input term to generate the term's score:

$$w_i(t) = \sum_{j=1\ldots L} log\left( \mathbb{1}(v_i = t_j)\left( ReLU(h_j W + b) \right) + 1 \right) \quad (1)$$

where $w_i$ is the *i-th* token in the vocabulary $\mathcal{V}$.

As described in Figure 1, an MLP is comprised of a Linear layer on top of a transformer encoder. The Linear layer takes the last hidden states of the transformer's encoder as input and projects each state to a positive scalar representing the weight of the corresponding input token.

An MLP encoder requires the input to be tokenized into a sequence of vocabulary words; therefore, it can only encode textual data and not images. In addition, MLP encoder does not have the capability to expand the input to relevant terms.

**MLM Encoder.** Unlike the MLP encoder, the MLM encoder can be applied to both text and image and has the freedom to expand the input to any relevant terms in the vocabulary. As described in Figure 1 (right), the MLM encoder consists of a transformer encoder and a sparse projection layer on top. The transformer encoder takes either a textual or visual input and outputs a single intermediate dense vector which is then passed into a sparse MLM projection:

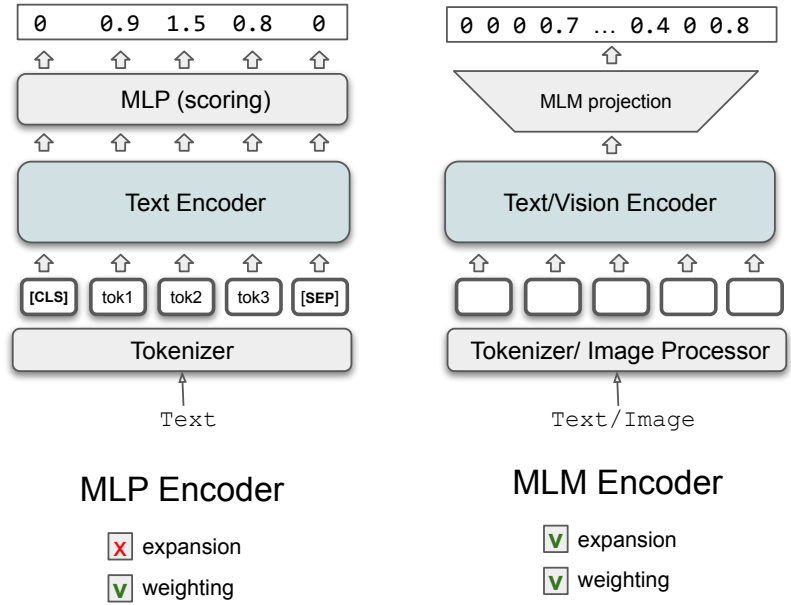

Figure 1: MLP and MLM encoder

$$w_i(t) = ReLU(h_0^\intercal e_i + b_i) \qquad (2)$$

where $w_i$ is the $i$-th item (token) in the vocabulary $\mathcal{V}$.

## 2.1 Full Bi-Encoder Configuration

The full bi-encoder configuration includes a query encoder, $f_\theta^Q$, and a document encoder, $f_\phi^\mathcal{D}$. Each query and document encoder is either a MLP or MLM as described in the previous section. In this paper, we experimented with different configurations as listed in Table 1. We did not explore all possible configurations due to resource constraints. Among all of the abovementioned configura-tions, the first model (M1) is the multi-modal version of Splade [7]. M2 is the multi-modal version of EPIC[26]. In M3 and M4, we re-use the MLP query encoder from M2 to encode the caption associated with each image.

| Model | Query encoder $f_\theta^Q$ | Document Encoder $f_\phi^\mathcal{D}$ | |
|---|---|---|---|
| | | Caption | Image |
| M1 | MLM | - | MLM |
| M2 | MLP | - | MLM |
| M3 | MLP | MLP (*) | - |
| M4 | MLP | MLP (*) | MLM |

Table 1: Sparse Bi-encoder Variants. (*) The MLP encoder is re-used from the query side for encoding caption.

## 3 EXPERIMENTS

**Dataset.** We conducted our experiments on the AToMiC dataset[45] which has around 11 million images and more than 10 million text queries collected from Wikipedia. Each text query contains four different fields, including *page_title*, *section_title*, *context_page_description*, *context_section_description*. We exclude the *context_page_description* and concatenate the remaining fields with space between them to form a single text query. Each image in the dataset also comes with a multilingual caption (*caption_reference_description*), we tested different configurations where we use either or both of them to represent the document. Due to computing resource limitations, our experiments were only conducted on images with English captions. For training, we only use the 4.4 text-image pairs provided by the dataset's authors and train for 5 epochs with InforNCE loss and only in-batch negatives.

**Metrics.** To evaluate model performance, we report NDCG@k, MAP@K, and R@k where $k = \{5, 10, 100, 500, 1000\}$.

## 4 DISCUSSION

One of our criteria for model selection is that the model should produce meaningful, interpretable sparse vectors (bags of terms). However, we found that model M1 does not meet this requirement. As demonstrated in Table 5, M1 generates terms that do not reflect the content of the image and are difficult for humans to interpret. The reason is that M1 uses the MLM encoder both on query and document sides, allowing the input text and image to be projected into any latent dimensions as long as these dimensions are co-activated (having non-zero values) in both query and document

Table 2: Performance of the submitted models on the image suggestion task, evaluated on NDCG@k scores.

| Run | NDCG@5 | NDCG@10 | NDCG@100 | NDCG@500 | NDCG@1000 |
|---|---|---|---|---|---|
| $M_{\mathcal{T}\to\mathcal{I}}$ | 0.40 | 0.39 | 0.28 | 0.50 | 0.58 |
| $M_{\mathcal{T}\to\mathcal{T}}$ | 9.39 | **10.79** | 15.59 | 18.67 | 19.59 |
| $M_{\mathcal{T}\to\mathcal{T}+\mathcal{I}}$ | **9.77** | 10.74 | **15.63** | **18.79** | **19.68** |

Table 3: Performance of the submitted models on the image suggestion task, evaluated on MAP@k scores.

| Run | MAP@5 | MAP@10 | MAP@100 | MAP@500 | MAP@1000 |
|---|---|---|---|---|---|
| $M_{\mathcal{T}\to\mathcal{I}}$ | 0.02 | 0.04 | 0.04 | 0.05 | 0.05 |
| $M_{\mathcal{T}\to\mathcal{T}}$ | 3.51 | **4.48** | 5.82 | 6.04 | 6.07 |
| $M_{\mathcal{T}\to\mathcal{T}+\mathcal{I}}$ | **3.58** | **4.48** | **5.84** | **6.07** | **6.10** |

Table 4: Performance of the submitted models on the image suggestion task, evaluated on Recall@k scores.

| Run | R@20 | R@100 | R@500 | R@1000 |
|---|---|---|---|---|
| $M_{\mathcal{T}\to\mathcal{I}}$ | 0.37 | 0.43 | 1.31 | 1.61 |
| $M_{\mathcal{T}\to\mathcal{T}}$ | **15.54** | 24.54 | 37.00 | 41.24 |
| $M_{\mathcal{T}\to\mathcal{T}+\mathcal{I}}$ | 15.50 | **24.57** | **37.48** | **41.37** |

Table 5: Demonstration of not interpretable output. Top-10 highest-scored terms are shown for both M1 and M2 models.

| Image | M2 (MLP, MLM) | M1 (MLM, MLM) |
|---|---|---|
| 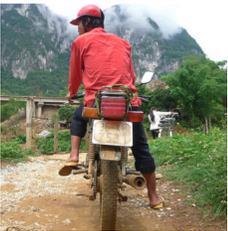 | mountain mountains bike bee dirt mo red path ##oot person | accent ship natural de crown yourself " ra now wild |

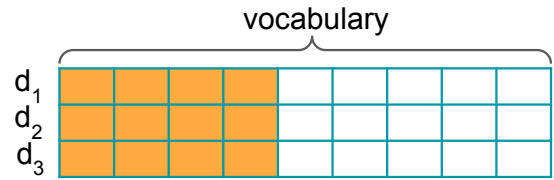

(a) Output dimensions are densely co-activated

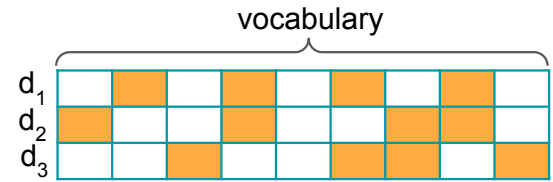

(b) Output dimensions are sparsely co-activated

Figure 2: Dense activation vs. sparse co-activation. All documents have 4 vocab terms activated (yellow colors).

representations. The projection freedom of MLM leads to the issue of high co-activation, forming a sub-dense space inside the vocabulary space as described in Figure 2. This high co-activation could also harm the retrieval efficiency of the inverted index employed in LSR. Because of the above issues, we did not submit the run from M1 for evaluation.

One solution to the above issues is to constrain the projection to semantically relevant terms in the vocabulary. This constraint could be achieved by using the MLP encoder on the query side. Indeed, the design of the MLP encoder only allows it to score the impact of the input tokens but does not allow it to expand to other tokens. MLP only produces positive weights for tokens in the input query and keeps the remaining tokens in the vocabulary to be zero. In order to match an image to a relevant query, the image encoder (MLM) needs to project the image into terms appearing in the relevant query. This constraint forces the model to produce more interpretable and sparsely co-activated image representation. However, we observe that using the MLP query encoder still does not prevent the problems entirely as it could still rely on stop words and punctuation marks for encoding latent senses. The stop words and punctuation marks are especially popular in long texts, as in the Atomic queries, which are taken from Wikipedia articles. For this reason, we resorted to using the short caption and image pairs for training our M2, M3, and M4 models. An example output of M2's document encoder-trained caption, image pairs is shown in Table 5.

Regarding the effectiveness, for each model (M2, M3, M4), we submitted one run for evaluation. The results are shown in Table 4, Table 3, and Table 2 for Recall, MAP, and NDCG respectively. We observe that M2 performs poorly across different metrics, implying that the task of image suggestion for writing assistants solely based on the images' content is a challenging task. We hypothesize that the context information of an image, which is critical for the task, could not be encoded in the images, but in text captions. Given a picture of World War 2 (WW2), for example, it is generally very difficult for a model to predict that this picture is about WW2 because any war picture has similar concepts (e.g., "soldiers", "weapon"). Similarly, given a picture of a less-popular street in Amsterdam, it is difficult to infer any terms relevant to Amsterdam based on the image content only. For this reason, we argue that to solve the image suggestion task well, the image caption should be used to provide more fine-grained and contextual information that is challenging to infer from the visual data. This argument is supported by the result of our second run produced by my M3 model. By using only image caption, the M3 model could outperform M1 significantly and by a large margin. The result of M4 also shows that using both images and captions could slightly improve the overall performance of the task, but the improvement is not consistent across different metrics.

## 5 RELATED WORK

**Learned sparse retrieval (LSR).** LSR is a neural retrieval method encoding queries and documents into sparse lexical vectors, efficiently indexed and searched with an inverted index. Various LSR approaches exist, using MLP or MLM encoders [7, 30, 47]. MLP encoders predict term importance without expansion, while MLM encoders use masked language model logits for weighting and expansion. Splade is a recent text-oriented LSR approach employing MLM encoders [6, 7], while other methods use MLP encoders [4, 21, 26]. Recent research suggests a cancellation effect between query and document expansion [31].

**Cross-Modal Retrieval (CMR).** CMR methods create a multi-modal representation space, measuring concept similarity across modalities [2]. Early CMR approaches used canonical correlation analysis [10, 14], followed by RNN-CNN encoders with hinge loss [8, 43], hard-negative mining [5] and attention mechanisms such as dual attention, stacked cross-attention, and bidirectional focal attention [16, 24, 28, 36]. Other approaches include modality-specific graphs [42] and image-text generation modules [11]. Domain-specific research targets CMR in fashion [9, 15], e-commerce [12], conversational systems [34, 35, 37], and music video recommendations [41].

Recent approaches leverage transformer-based dual encoders pretrained on large-scale datasets, e.g., with the alignment of uni-modal representations before fusion [18] and usage of adaptation models for learning representations on the level of objects [46]. Zeng et al. [48] propose to deploy an additional cross-modal encoder for refining VL representations, while Yuan et al. [46] use adaption models for learning representations on the level of objects. Other approaches imply large-scale contrastive pretraining on image-caption [32] and alt-text pairs [18]. Related to this, another line of work adopts BERT-based encoders [40] for the task [27], including both two-stream [25, 39] and single-stream architectures [1, 3, 17, 19, 38]. Related to this, Li et al. [20] propose to incorporate caption object tags with region features, while Wang et al. [44] propose to adapt multiway transformers.

**Our focus.** Unlike prior work that focuses on sparse to dense conversion [22, 23], we focus on dense to sparse conversion in the multi-modal domain. Challenges include dimension co-activation and semantic deviation [33].

## 6 CONCLUSION

In this work, we explored the application of learned sparse retrieval (LSR) for the image suggestion task to support multimedia content creation. We identify the challenges that arise when transferring state-of-the-art LSR techniques from the text domain to the multi-modal domain and propose a simple solution to mitigate the problems. We analyze the effectiveness of our trained models with various configurations and conclude that using image captions is critical for the task as image captions provide fine-grained concepts and context information that are difficult to encode in the visual content itself. We address this problem in the follow-up work [29].